\documentclass[twocolumn,showpacs,preprintnumbers,amsmath,amssymb]{revtex4}
\usepackage{graphicx}% Include figure files
\usepackage{dcolumn}% Align table columns on decimal point
\usepackage{bm}% bold math

\begin{document}

\title{ The superheated Melting of Grain Boundary }
\author{Wei Fan}
\email{fan@theory.issp.ac.cn}
 \affiliation{ Key Laboratory of
 Materials Physics, Institute of Solid State Physics, Chinese
 Academy of Sciences, 230031-Hefei, P. R. China}
\author{Xin-Gao Gong}
\affiliation{Department of Physics, Fudan University, 200433-Shanghai, P. R. China \\
 Key Laboratory of Materials Physics, Institute of Solid State Physics, Chinese
 Academy of Sciences, 230031-Hefei, P. R. China}

\begin{abstract}
 Based on a model of the melting of Grain Boundary (GB), we
 discuss the possibility of the existence of superheated GB state.
 A Molecular Dynamics simulation presented here shows that the
 superheated GB state can realized in the high symmetric tilt GB.
 Whether the sizes of liquid nuclei exceed a critical size
 determined the superheating grain boundary melting or not. Our
 results also indicate that the increase of melting point due to
 pressure is smaller than the superheating due to nucleation
 mechanism.
\end{abstract}


\maketitle

\section{\label{sect1} Introduction }

 The superheating has been found in a larger number of systems such as
 surface~\cite{Carnevali,Pluis,Denier} , small cluster~\cite{SHVART}, confined
 thin film~\cite{ZHANGL1} and particles covered (or embedded in) by material with
 higher melting point~\cite{Daeges}. Generally, the melting of solid material is
 heterogeneous process with the nucleation mechanism at surfaces or
 interfaces~\cite{Cahn,Madd}.  Providing heterogeneous
 nucleation could be avoided by means of suitable
 coating~\cite{Daeges} or internal heating~\cite{Khaikin}, the
 metal crystal can be in the superheated state, its melting is
 completed by a thermodynamically instability resulting in
 homogeneous disordering and catastrophic mechanism with the
 stability limit from 0.2$T_{m}$ to
 2.0$T_{m}$~\cite{Lele,Kauzmann,Fecht,Tallon,Lu}.

 A large number of researches have been contributed to the role of
 surface for the melting of crystal.  The superheated melting
 of fcc(110) and fcc(100) surfaces are virtually never
 observed~\cite{Frenken,Tolla,Sun}. The only example of the
 superheated surface is the small crystal strictly confined by
 high-symmetry fcc(111) facts~\cite{Carnevali,Pluis,Denier}. GB as
 another important quasi-2D defect also leads to the
 heterogeneous melting of solid material.  Quite a number of
 studies have shown that GB can't melt below the $T_{m}$ (not
 premelting)
 ~\cite{Nguyen,Kikuchi,Ciccotti,Nguyen1,Lutsko,Lutsko1,Broughton,Plimpton,Carrion,Plimpton1,Fan,Phillpot}.
 Using the MD simulation, Kikuchi and Cahn, Ciccotti $et.  al.$
 showed that GB doesn't melt until temperature reaches to melting
 point of bulk.  Nguyen $et.  al.$ using the more accurate
 interatomic potential by embedded-atom methods(EAM) studied the
 high-temperature GB structure, they found that,
 close to melting point T$_{m}$, the GB structure was disordered, quite
 liquid-like and meta-stable, and over a long interval of simulation
 the underlying crystalline order can re-emerge.  The experiment
 (T.E.Hsieh and R.  B.  Balluffi) using the HREM(High Resolution
 Electron Microscopy) methods supported above arguments and showed
 that aluminum GB did not melt below 0.999$T_{m}$~\cite{Hsieh}.

 It is more surprising that some the simulations also hint that
 some high symmetric GBs similar to high symmetric surface can
 probably melt by the superheated~\cite{Broughton}.  In this work,
 parallel to Di Tolla's work~\cite{Tolla} for surface we study the possibility of
 superheated high-symmetry GB by a theoretical model and MD simulation
 of a symmetric aluminum GB.

 When temperature beyond melting point a crystal reaches the superheated state, all
 liquid nuclei must be smaller than a critical size. These liquid nuclei are unstable and
 able to re-crystallize again. The liquid nuclei easily form at surfaces,
 grain boundary and other solid defects regions. So, to avoid larger liquid nucleus, the
 crystal must be prepared with the lowest number of solid defects.
 It's easier to study the superheating of crystal in computer
 simulation than in experiment. We can construct prefect crystal
 using the periodic boundary condition in computer simulation. In
 experiment it's very difficult to obtain infinite volume prefect
 crystal. We can't eliminate the influence of surfaces,
 grain boundaries, dislocations and other defects with complicated
 structures. All these defects have potentially become the
 liquid nuclei to melt crystal. However surface effects can partially remove by
 coating other material with higher melting point or internally heating the material.
 By these methods we can obtain superheated crystalline grains.

 Additionally, the grain boundary itself may probably become the liquid nucleus near
 the melting point of crystal. Based on our theoretical model,
 it's possible that, under condition of the absent of critical nucleating cores,
 grain boundary doesn't melt even the temperature beyond melting point.
 We hope to find the superheated grain boundary in computer simulation, although
 it is difficult to find in experiments due to different kind of
 unavoidable nucleation mechanism. Our simulation will show that high-symmetry
 grain boundary can sustain above the melting point of crystal without
 other nucleation mechanism. The superheated grain boundary is
 easily understood by proximity effects if we consider the grain boundary
 is sandwiched between two superheated crystalline grains, while the
 superheated grains can be obtained by properly internal heating.

 \begin{figure}
 \resizebox{7cm}{13cm}{\includegraphics{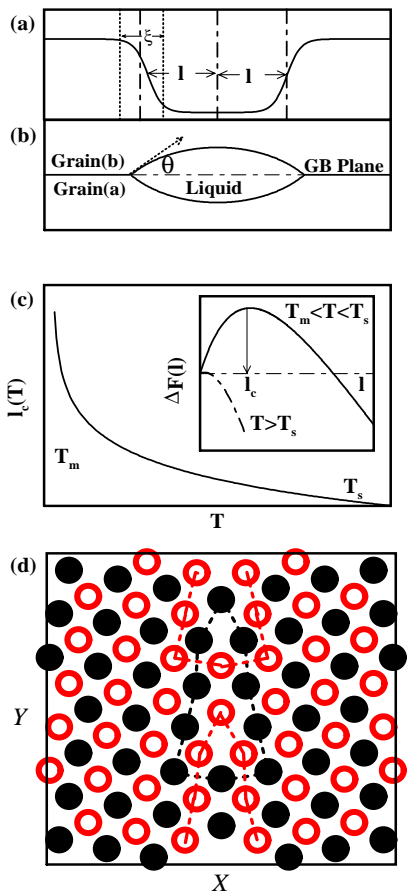}}
 \caption{\label{fig1} (a) The change of atomic density across the
 melting GB. (b) The schematic of partly wetting GB.  (c) Critical
 liquid thickness of a non-melting GB vs temperature above T$_{m}$
 (schematic). Inset:  free energy change upon conversion of a film
 of thickness $l$ from solid to liquid.  From $T_{m}$ to $T_{s}$
 the solid GB is stable. (d)The structure Al $\Sigma$={13} GB. The
 solid and open circles show the I and II(001) planes of the ...I
 II I II..., stacking sequence along z. The basic GB structural
 units are also shown with dash-dot lines. This figure illustrates
 only part of region near grain-boundary plane of our simulation cell}
 \end{figure}

\section{\label{sect2}  The Model of GB Melting}

 The melting point $T_{m}$ of a solid may be defined as the
 temperature with the coexistence of solid phase and liquid phase.
 For a solid with surfaces or grain boundaries, the melting is
 generally completed by the mechanism of heterogeneous nucleation.
 We consider that a liquid film with thickness $2l$ forms between
 two semi-infinite solid (Fig.~\ref{fig1}(a)).  The change of free
 energy per unit area is taken as

 \begin{equation}
  \Delta F(l)= 2\rho L l (1-T/T_{m}) + \Delta \gamma (l)
 \label{dfl}
 \end{equation}

 \noindent where $\rho$ is the liquid density, $L$ the latent heat of
 melting, $\Delta \gamma(l)$ the difference between the overall free
 energy $\gamma_{SL-SL}$ of two interacting solid-liquid interfaces
 separated by a distance $l$ and the GB energy per unit area

 \begin{equation}
  \Delta \gamma(l) = \gamma_{SL-SL} - \gamma_{GB}
  \label{dgl}
 \end{equation}

 By extending the Cahn's wetting theory to solid-solid
 interface~\cite{Gennes}, using $\Delta\gamma(0)$ = 0 and only
 considering the short range interaction, we may obtain

 \begin{equation} \Delta \gamma(l) =
 \Delta\gamma_{\infty}(1-e^{-l/\xi}).
 \label{dgl1}
 \end{equation}

 \noindent where $ \Delta\gamma_{\infty}=2\gamma_{SL} - \gamma_{GB}$ is
 the difference of the interface energy of two isolated solid-liquid
 interface $\gamma_{SL}$ and the GB energy $\gamma_{GB}$, $\xi$ is the
 width of solid-liquid interface.

 The condition of GB melting is defined by following process.  We
 may image there is a droplet in GB region (Fig.~\ref{fig1}(b)),
 the equilibrium condition of the droplet is

 \begin{equation}
  \gamma_{GB}-2\gamma_{SL}cos(\theta)=0
  \label{ggb}
 \end{equation}

 \noindent where $\theta$ is the wetting angle. If
 $\Delta\gamma_{\infty}>0$, $i.e.$ $\gamma_{GB}<2\gamma_{SL}$,
 $\theta$ is a finite value, the droplet can survive in the region
 of GB (partial wetting), the whole GB can't be wetted.  As
 $\Delta\gamma_{\infty}<0$, $i.e.$ ,$\gamma_{GB}>2\gamma_{SL}$,
 $\theta$ can't be defined. A liquid film forms in GB region
 (Wetting) and the GB melts.  Thus $\Delta\gamma_{\infty}=0$, $i.e.$
 $\gamma_{GB}=2\gamma_{SL}$ and $\theta=0$, may be considered as
 the criteria of GB Melting.

 In early time, one was decline to think that GBs melt below melting
 temperature, that is, at certain temperature below $T_{m}$,
 $\Delta\gamma_{\infty}<0$.  In
 this paper we only consider the possibility of GB superheated melting,
 that is, as $T>T_{m}$, $\Delta\gamma_{\infty}>0$.

 For superheated melting, the insert of Fig.~\ref{fig1}(c) show
 that $\Delta F(l)$ has a local minimum at $l=0$, and approaches to
 negative infinite as $l$ approach infinite.  There is a maximum at
 $l_{c}$, which is

 \begin{equation}
  l_{c}= \xi Ln(\frac{\Delta\gamma_{\infty}
  T_{m}}{2L\rho\xi (T-T_{m})})
  \label{lc}
 \end{equation}

 The Fig.~\ref{fig1}(c) shows the temperature dependence of $l_{c}$.
 According to Fig.~\ref{fig1}(c), at a certain temperature
 $T(>T_{m}$) and $l<l_{c}$, the system can reduce the its free
 energy by decreasing the thickness of liquid film until the
 thickness reaches to zero, that is, the system crystallizes.  When
 $l>l_{c}$, the system can reduce its free energy by increasing the
 thickness of liquid film until the thickness reach to infinite,
 the GB melts.  $l_{c}$ is the critical thickness at $T$. If at a
 temperature $l_{c}=0$, any small thickness can lead to the melting
 of GB.  The temperature is named as the maximum superheated
 temperature $T_{s}$, expressed as

 \begin{equation} T_{s} =
 T_{m}(1+\frac{\Delta\gamma_{\infty}}{2L\rho\xi})
 \label{ts1}
 \end{equation}

 \noindent $l_{c}$ is also be expressed as

 \begin{equation} l_{c} = \xi Ln(\frac{T_{s}-T_{m}}{T-T_{m}})
 \label{lc2}
 \end{equation}

 The critical nucleus is extremely large for $T \approx T_{m}$
 and reduces rapidly with the increase of the superheating degree.
 At $T_{s}$, $l_{c}=0$ (Fig.~\ref{fig1}(c)) and the
 spontaneous melting happens. Between $T_{m}$ and $T_{s}$ the
 superheated states is meta-stable, although the melting doesn't
 occur.  In following several sections, by MD simulations of
 Aluminum GB melting and above model, we prove that GB can
 preserve crystalline even above $T_{m}$ until temperature reaches to
 the maximum superheated temperature $T_{s}$.  Our simulation shows
 the behavior of superheating GB, that is, for $T_{m}<$ $T$
 $<T_{s}$, there exists a critical width $l_{c}$ of liquid film.
 When the width of the artificially added liquid is larger than
 $l_{c}$ the GB melts, or the liquid film will reduce and the
 effect of crystallization is dominating.

 \begin{figure}
 \resizebox{8cm}{11.0cm}{\includegraphics{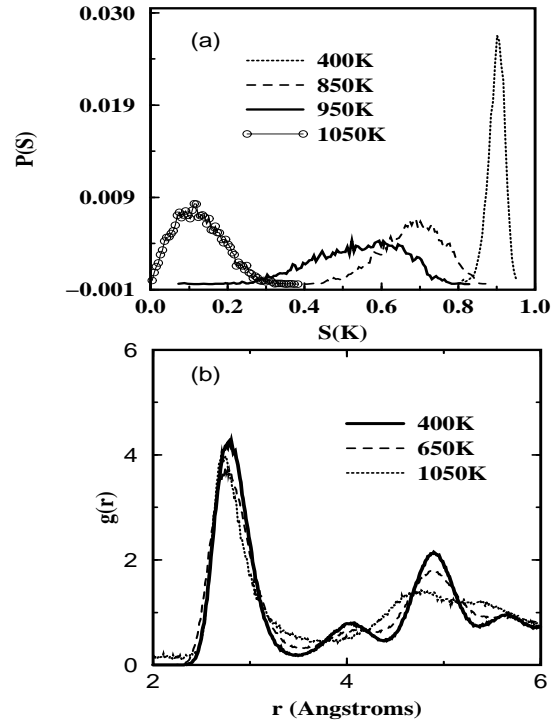}}
 \caption{ \label{fig2} The distribution functions $P(S)$ (a) and
 the pair correlation functions (b) at various temperatures.  The
 flat and wide distribution of $S({\bf K})$ at 950K represents the
 coexistence of liquid and solid.  }
 \end{figure}

\section{\label{sect3}  The Molecular Simulation of Gain Boundary}

 Molecular Dynamic simulations of crystal~\cite{JIN} and high-symmetry
 surface~\cite{Tolla} have shown the superheating crystal without
 critical nuclei such as point defects and liquid drops.
 By internal heating, crystal with high-symmetry fcc(111) surface
 will be superheating. This is because the process of nucleation is homogeneous in
 the surface region. For some low symmetry surfaces such as fcc(110) surface, the
 anisotropy leads to heterogeneous nucleation and high concentration of defects,
 superheating is extremely difficult to achieve.

 Reconstruction and roughening are two main structural transitions for a surface when
 increasing temperature. Behavior of grain boundary is more complicated than surface,
 including the migration, bending, sliding, zigzag and faceting transition. Our
 simulations show that it is difficult to control the homogeneous nucleation in computer
 simulation to obtain superheated grain boundary. However we can use symmetric grain
 boundary to make atoms homogenously distribute in GB region. Instead of periodic
 boundary, we have fixed two boundaries of simulation cell parallel to GB plane to
 decrease the possibility of grain-boundary sliding.

% \begin{figure}
% \resizebox{9cm}{8cm}{\includegraphics{fig3}}
% \caption{\label{fig3} The configuration of GB at high temperature.
% (a) The GB structure at 850K (high temperature disorder state) (b)
% The GB structure at 1050K (melting state) (c,d) The GB structure
% at 975K (superheated state) with unstable liquid film. }
% \end{figure}
 \begin{figure*}
 \includegraphics{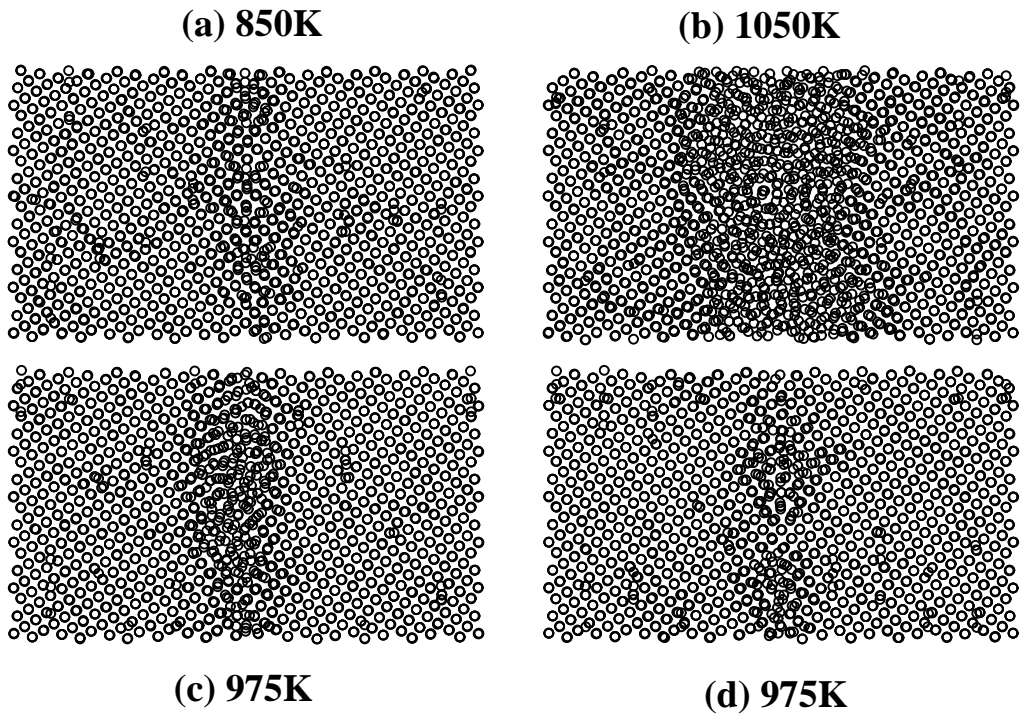}
 \caption{\label{fig3} The configuration of GB at high temperature.
 (a) The GB structure at 850K (high temperature disorder state) (b)
 The GB structure at 1050K (melting state) (c,d) The GB structure
 at 975K (superheated state) with unstable liquid film. }
 \end{figure*}

 In this work we choose Al symmetric $\Sigma$13 (320) [001] tilt
 grain boundary Fig~\ref{fig1}(d) as our simulation cell with
 mis-orientation $67.8^{\circ}$, the tilt axis of the boundary
 is along the Z direction, the fixed boundary-conditions is used
 in the X direction perpendicular to GB plane, and periodic boundary
 condition are used in the Y and Z directions. The widths along X, Y, Z
 are 100$\AA$, 43$\AA$, 12$\AA$ respectively. Interatomic potential
 plays the most important role in molecular dynamics simulations.
 In this paper, by using a more realistic potential, a Glue
 potential developed by F.  Ercolessi and J.  B. Adams~\cite{E1},
 The glue potential has been used in a large number of the
 simulations of surface, cluster, liquid, and crystal, and the
 simulation results are perfectly consistent with experimental
 results~\cite{Carnevali,Tolla,Shu,E3,E4}. The lattice constant
 $a_{0}$ for Al at 0K is 4.032$\AA$. For the glue potential the
 melting points is about 936K~\cite{E2}, which is close to the
 experimental melting point of Aluminum (about 933K). The MD
 simulations are carried out at constant temperature, constant
 volume and constant atomic number.  The time step is 0.06 ( about
 0.003psec ), at each temperature the runs are made about 20000
 time steps ( about 60psec ).

 \begin{figure*}
 \includegraphics{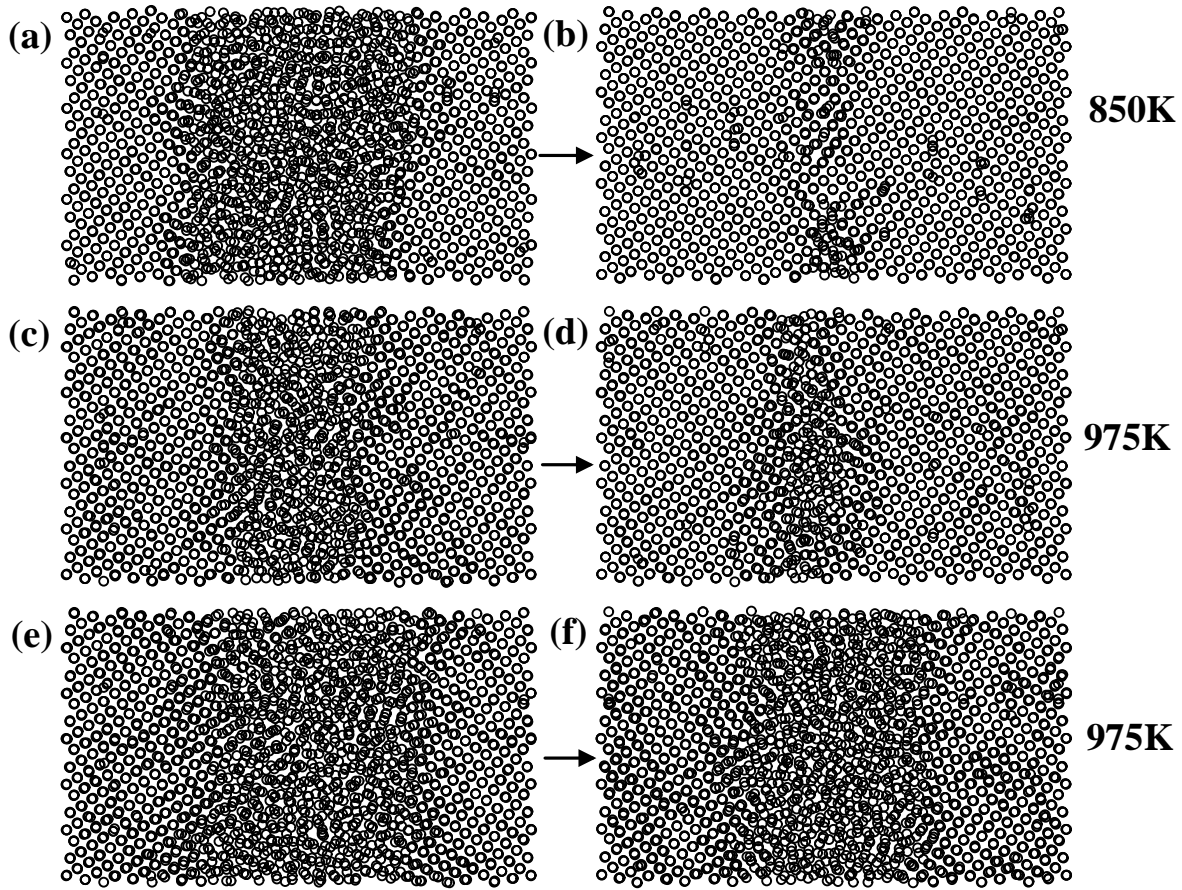}
 \caption{\label{fig4} The response of superheated state for adding
 liquid film.  (a,b) at 850K, $2l=$30$\AA$. (c,d) at 975K,
 $2l=$20$\AA$. (e,f) at 975K, $2l=$30$\AA$. }
 \end{figure*}

 The static structure-factor $S({\bf K})$ for specific reciprocal-space vector
 ${\bf K}$ (Eq.~\ref{sk1}) represents a quantitative measure for the long-range order
  and can be used as order parameter describing the transition between disorder and order.

 \begin{equation} S({\bf K}) = < \mid \sum_{i\in GB}
 \exp{(i{\bf K}\cdot {\bf r}_{i}(t))} \mid^{2} >/N_{GB}^{2}
 \label{sk1}
 \end{equation}

 \noindent ${\bf r}_{i}(t)$ is the position of $i$th atom at time
 $t$. $<\cdots>$ is indicative of the time average. $N_{GB}$ is
 the number of atoms in grain-boundary region.  For a crystal
 the $S({\bf K})$ is approaching 1, for liquid it approaching to 0.
 In order to study the stability and the nature of disordered
 grain-boundary, we define a distribution function of $P(S)$ which
 is the statistics of the value of $S({\bf K})$ (${\bf
 K}=\frac{2\pi}{a_{0}}(0,0,1))$ of all molecular-dynamics time
 steps.  Fig.~\ref{fig2} shows the distribution functions at
 several temperatures from 400K to 1050K.
 The centers of the peaks represent the degree of disorder and
 order, the widths of the peaks as a measure of the fluctuation of
 $S({\bf K})$ represent the instability of the GB structure.
 From Fig.~\ref{fig2}, the peak is very narrow and the
 center of the peak is very near 1 in low temperature regime
 ($T\leq$ 400K) , and this implies that the structure of grain
 boundary is very order and stable in low temperature regime; At
 850K, the peak is board ( the minimum is at about 0.5 and the
 maximum is at about 0.9 ) and the center of the peak is much less
 than 1 , which implies that the GB structure is disordered in this
 temperature.  At about 950K, the width of peak is extremely board
 (the minimum reaches 0.1 and the maximum still retains at about
 0.9), the fact implies the structure of grain boundary is rather
 unstable, sometime the structure of grain boundary is rather
 disorder like liquid because of the $S({\bf K})$ approaching 0, and
 sometimes it just likes a crystal structure with long range order
 because of the $S({\bf K})$ probably approaching 1.  This shows
 that at this temperature the coexistence of liquid phase and solid phase is reached.
 Above results show that the melting-point of this system is very close to 950K.
 In our simulation, the large $S({\bf K})$ fluctuation near 950K just indicates the signal of
 solid-liquid phase transition. At 1050K the peak moves to the left and become narrow. The
 position of the peak is close to zero, the GB is melting. Our results also show that the maximum
 superheated temperature $T_{s}$ is close to 1050K. We also
 calculated the pair correlation functions $g(r)=\frac{1}{4\pi\rho
 N}<\sum_{i\ne j}\delta(r-r_{ij})>$ at various temperatures which
 show the liquid behavior at 1050K. Fig.~\ref{fig3}(a,b)
 illustrate the GB structures at 850K and 1050K.

 \begin{figure}
 \resizebox{8cm}{8cm}{\includegraphics{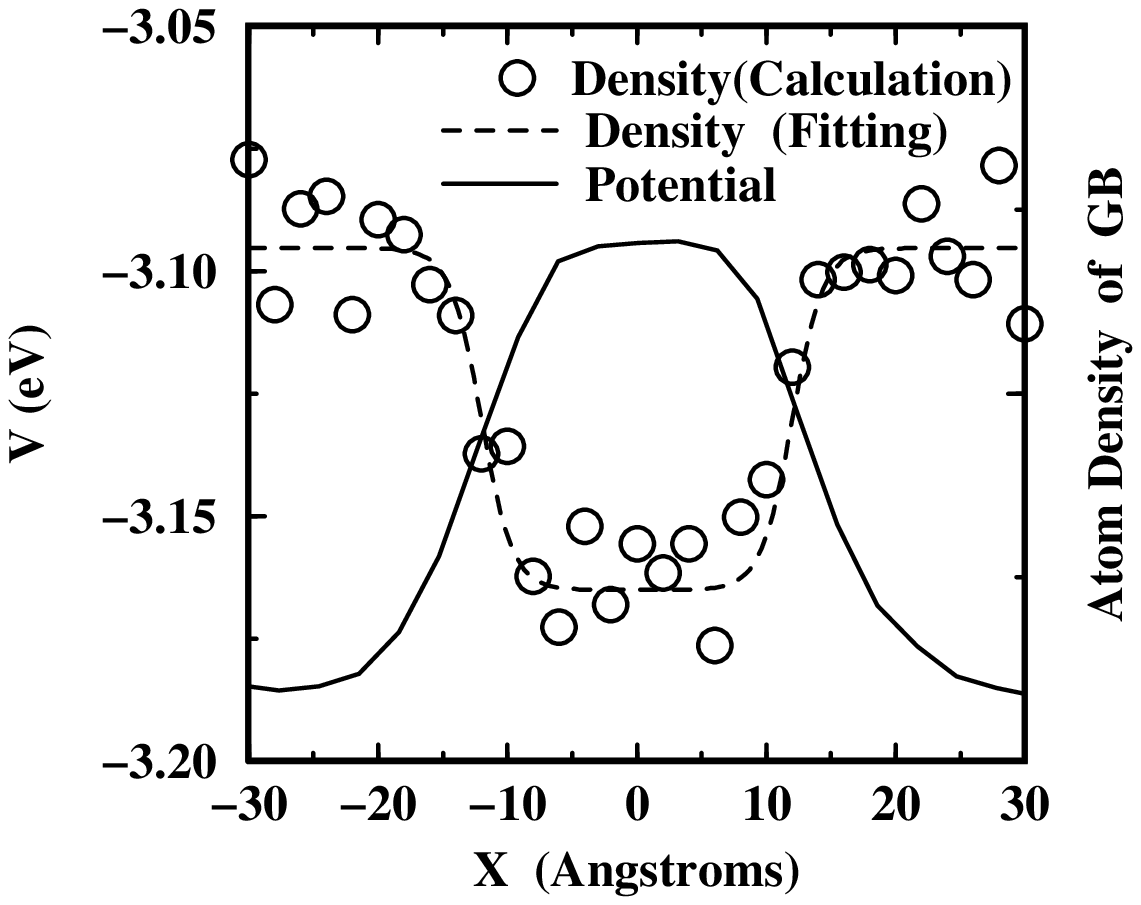}}
 \caption{\label{fig5} The profile of potential and atomic density
 across the GB at 975(K). }
 \end{figure}

 Our model shows that between $T_{m}$ and $T_{s}$, GB will enter a
 new superheated state of GB.  The new state is characterized by
 (1) the coexistence of liquid and solid; (2) the smaller size of liquid
 nucleus than that of the critical nucleus at that temperature prevents
 from the melting of GB although $T > T_{m}$. Fig.~\ref{fig3}(c,d) show the competition of
 liquid and solid phase at 975K in superheated state.  Sometimes there exists
 a liquid-like layer in the grain-boundary region (c) but it is meta-stable and may
 disappear and crystalline phase re-emerges at following time steps (d). We
 will show that the superheated state (975K) is rather different
 from the high-temperature disordered state at 850K.

 A liquid layer as the melting nucleus can be artificially
 added to the GB region by following methods:  At a certain
 temperature $T$, we sample some layers with width $2l$, the atoms
 in these layers are heated up to an appropriate temperature
 $T_{a}$($>T$) until a liquid layer forms.  By allowing all atoms
 relaxation at temperature $T$ again, we can obtain a new
 equilibrium structure at temperature $T$. Fig.~\ref{fig4} shows
 that both the initial GB configurations Fig.~\ref{fig4}(a,c,e)
 having already added a liquid film with widths $2l$ and the
 final equilibrium configurations Fig.~\ref{fig4}(b,d,f) at 975K and
 850K respectively.  At 850K, for $2l=30\AA$, the liquid layers
 disappears after the relaxation about 50ps Fig.~\ref{fig4}(a,b).
 However at 975K for $2l=20\AA$, the liquid
 layer disappears Fig.~\ref{fig4}(c,d)and for $2l=30\AA$ the layer of liquid is still
 existent Fig.~\ref{fig4}(e,f) after relaxation about 50ps.
 Therefore, we can obtain $20\AA$ $<2l_{c}<$ $30\AA$ at $T$=975K.

 Above results also show that the superheated state (975K) is very different in nature
 from the high temperature disorder state (850K).  For high temperature
 disorder state, the liquid film can't induce the melting of GB,
 but for superheated GB state, the GB melts only when the width of liquid film is larger
 than a critical width $l_{c}$. In order to define the correlation length
 $\xi$ and the thickness of liquid film, we calculate the atom
 density profile corresponding Fig.~\ref{fig4}(f). Our results
 show that at 975K, the critical width $2l_{c}$ is between $20\AA$
 and $30\AA$ and $\xi=10\AA$, thus $1.0<l_{c}/\xi<1.5$, which is
 consistent with the theory model $l_{c}/\xi=1.4$ with $T_{m}=950K$
 and $T_{s}=1050K$ in Eq.~\ref{lc2}.

\section{\label{sect4} Discussion and Conclusion}

 In summary, we study the possibility of superheated GB state by
 both a theoretical model and MD simulation. Our results indicate that we can
 obtain superheated grain boundary by having properly controlled homogeneously
 nucleating precession when increasing temperature. If there are liquid nuclei
 whose sizes are larger than a critical size the superheated grain boundary melts.
 Or the grain boundary waits for homogeneously melting when
 temperature higher than maximum superheated temperature.

 We must justify that pressure plays important roles for the
 superheated in experiments and computer simulations. In
 experiments, both coating with high-melting-point materials and
 heating internally induce the internal pressure in melting region. In
 our simulation also there is internal pressure in melting region
 because the size of simulation cell doesn't change companying with the
 increasing temperature. In this paper we only consider the
 superheating state due to the nucleation mechanism. The pressure
 mechanism and nucleation mechanism become intertwined and
 influence the melting of materials. Pressure lead to the
 increase of melting points. The melting point is about 950K in
 our simulation and higher than the experimental melting points
 T$_{exp}$=933K and T$_{glue}$=936K~\cite{E2} in simulation using glue
 potential without internal pressure. Pressure increases melting
 point less than T$_{m}$-T$_{glue}\sim$950K-936K=14K. However the
 nucleation mechanism leads to the superheating about T$_{s}$-T$_{m}\sim$100K.

 The proximity effects of the superheated grains are also important to induce the
 superheated grain boundary sandwiched between two properly superheating grains.
 Because superheated state is meta-stable state in phase diagram,
 we don't expect it's long-live. The superheated materials melt by
 the nucleation mechanism or homogeneously melt at higher temperature.

\section*{Acknowledgements}

 Author is greatly indebted to professor D.Y.Sun and Yizhen He for
 valuable suggestions. This work is financially supported by Laboratory of
 Internal Friction and Defects in Solids, Chinese
 Academy of Sciences (currently named as Key Laboratory of
 Materials Physics, Chinese Academy of Sciences), the National
 Nature Science Foundation of China and Chinese Academy of Sciences
 under KJCX2-SW-W11.

\end{document}